\documentclass[12pt]{article}
\usepackage{graphicx}
\usepackage{lscape}
\begin{document}

\baselineskip=22pt

\title{A generalized model of active media with a set of interacting pacemakers:
Application to the heart beat analysis}

\author{SERGEI RYBALKO$^\dag$ and EKATERINA ZHUCHKOVA$^\ddag$}
\date{}
\maketitle

\begin{center}

{\small $^\dag$ Department of Physics, Kyoto University, Kyoto 606-8502, Japan}

{\small $^\ddag$ Physics Faculty, Moscow State University, Leninskie gory, Moscow
119992, Russia}

E-mail: rybalko@polly.phys.msu.ru

\end{center}

\begin{abstract}
We propose a quite general model of active media by consideration of the
interaction between pacemakers via their phase response curves. This model
describes a network of pulse oscillators coupled by their response to the
internal depolarization of mutual stimulations.

First, a macroscopic level corresponding to an arbitrary large number of
oscillatory elements coupled globally is considered. As a specific and important
case of the proposed model, the bidirectional interaction of two cardiac nodes is
described. This case is generalized by means of an additional pacemaker, which
can be expounded as an external stimulater. The behavior of such a system is
analyzed. Second, the microscopic level corresponding to the representation of
cardiac nodes by one-- and two--dimensional lattices of pulse oscillators coupled
via the nearest neighbors is described. The model is a universal one in the sense
that on its basis one can easily construct discrete distributed media of active
elements, which interact via phase response curves.
\end{abstract}



\vspace{1cm} {\bf Short title:} A generalized model of active media

\newpage

\section{Introduction}

Representation of an active distributed system by ensembles of coupled excitable
or oscillatory elements is very useful method of the analysis because it allows
to understand main dynamical processes inherent in the considered medium. As is
known, this approach goes back to the model of Wiener and Rosenblueth [Wiener \&
Rosenblueth, 1946], according to which a medium consists of single elements being
in one of three possible states: excited, refractory or rest. Later such models
as coupled limit cycle oscillators and chaotic maps [Kaneko, 1990; Shibata \&
Kaneko, 1998] have played an important role not only in a quite realistic
description of active media but also in the understanding of a possible behavior
of systems far from equilibrium. Many useful concepts like phase--locked
patterns, synchronization and spatio--temporal chaos have become popular due to
detailed studies of similar nonlinear models [Kuramoto, 1984; Kuramoto, 1995;
Winfree, 2000].

Investigations of such an example of an active medium as cardiac tissue are of
significant scientific interest owing to vital importance of its rhythm
stability. Real heart cells exhibit oscillatory properties (can be reset and
entrained), they are excitable and have a refractory time, during which they do
not respond to external stimulation. Hence, the heart can be considered as
consisting of oscillatory (conductive cardiomyocytes, which have automaticity)
and excitable (contractile heart cells, which do not initiate electrical activity
under normal conditions) elements.

Due to extraordinary complexity of the heart, many qualitative discrete and
continuous models have been tested. Majority of computational models of cardiac
tissue of last generation takes into consideration the kinetics of excitable
cells, how the excitation propagates from cell to cell and how contractile
cardiomyocytes are arranged and connected in space [Clayton, 2001]. Such models
mainly serve for studying sustained by re-entrant activity lethal arrhythmia --
ventricular fibrillation, during which the spatio--temporal behavior is very
complex [Clayton {\it et al.}, 2006].

Other models treat the cardiac tissue as an active conductive system, taking into
account oscillatory properties of heart cells. In this case the cardiac rhythms
can be described on the basis of the dynamical system theory (see e.g.
[Courtemanche {\it et al.}, 1989; Goldberger, 1990; Bub \& Glass, 1994; Glass
{\it et al.}, 2002; Loskutov {\it et al.}, 2004] and refs. therein). Hereinafter
we hold this approach.

Under normal conditions the electrical activity of the heart (action potentials)
is spontaneously initiated in a region of the right atrium, sino--atrial (SA)
node, so--called leading pacemaker. Automatic excitation is a distinctive feature
not only of the cells of the SA node, but also of other conductive heart cells,
so-called latent pacemakers. In addition, contractile cardiomyocytes can initiate
a spontaneous action potential in pathology. Electrophysiological studies have
suggested that the activity of cardiomyocytes with automaticity (e.g. P-cells of
the SA node, of the atrioventricular (AV) junction, Purkinje cells) can be
modulated by current pulses stimulating (super-threshold depolarizing) applied
extracellularly [Jalife \& Moe, 1976; Sano {\it et al.}, 1978; Jalife {\it et
al.}, 1980; Antzelevitch {\it et al.}, 1982].

Effects of external stimuli on biological oscillators are observed in a wide
range of species. Experimentally obtained characteristics can be represented by a
phase response curve (PRC) [Jalife \& Moe, 1976; Antzelevitch {\it et al.}, 1982;
Reiner \& Antzelevitch, 1985]. To establish the shape of the PRC experimentally,
stimulation of an oscillator at various phases of its intrinsic cycle is applied
[Jalife {\it et al.}, 1980; Guevara \& Shrier, 1987]. It has been found that in
different pacemaker cells early stimuli delay the next pacemaker discharge and
late pulses advance it. Therefore, the typical PRC shape is biphasic [Jalife {\it
et al.}, 1980; Reiner \& Antzelevitch, 1985].

The rhythm of autonomous biological oscillators can also undergo an external
periodic perturbation (e.g. activity of cells of the AV junction is subjected to
sinus rhythm), depending on both the stimulus magnitude and its phase within the
cycle. It is known that when the frequency and the amplitude of the external
periodic stimulation are varied, a diversity of phase diagrams can be established
between the stimulus and the self-sustained oscillator (see e.g. [Loskutov {\it
et al.}, 2004]). In some situations the rhythm of the biological oscillator is
entrained (or phase-locked) to the external stimulation so that for each $M$
cycles of the stimulation there are $N$ cycles of the autonomous oscillator
rhythm. This occurs at a fixed phase (or phases) of the stimulus and is called
$M:N$ phase-locking or entrainment, which appears as a time--periodic sequence.
In particular, entrainment of $1:1$, during which the rhythms of the oscillator
and external stimulus are matched, is defined as synchronization phenomenon.

In the present paper we develop a general simplified model describing a network
of pulse oscillators coupled by their response to the internal depolarization of
mutual stimulations. Our primary aim is to keep the model as simple as possible
and to introduce a minimal number of parameters. Therefore, in our model the
pacemakers are fully characterized by their intrinsic cycle length and are
represented as pulse oscillators. Their interaction is described by PRCs. At
first, we will consider two bidirectionally interacting pacemakers to demonstrate
the basic concepts of the model. Then we will apply this approach to construct a
pacemaker network model with global coupling. As the following step, we will
analyze two specific cases of this PRC based model of coupled pulse oscillators:
two and then three interacting cardiac nodes. An additional pacemaker can also be
expounded as an external stimulater. Our further intention is to go on to the
next (microscopic) level and represent each pacemaker as an ensemble of
interacting oscillatory elements. Extrapolation of our approach to one-- and
two--dimensional matrixes (or lattices) of pacemaker cells interacting via
nearest neighbors concludes the present study.

\section{Development of the General Model}

In this part we construct a system with two pacemakers and then consider a quite
general model of a set of interacting pacemakers coupled by their PRCs.

\subsection{Two interacting pacemakers: outline of the approach}

Consider two interacting pulse oscillators (or pacemakers) $A$ and $B$ with
intrinsic periods of their autonomous beating $T_a$ and $T_b$ respectively. An
interaction between oscillators is governed by so-called phase response curve
(PRC). This means that a phase shift of one of the oscillators happens as a
result of an impact of the another one. To construct an adequate mathematical
model, it is necessary to accept some restrictions concerning the character of
the interaction. We describe them briefly [Ikeda, 1982; Glass {\it et al.}, 1986;
Glass \& Zeng, 1990].

1.  The phase of the disturbed oscillator is shifted to a new value instantly
after an impact.

2. The phase shift depends only on two main parameters: a) on the phase
difference of oscillators and b) on the influence strength. In turn, this
influence strength depends on its amplitude and the coupling coefficient of the
oscillators. In a real system the coupling coefficient is the average factor that
shows how the strength of the pulse decreases during its passing from one
oscillator to another. Thus, the phase shift $\Phi$ determining a new phase of
the disturbed oscillator with period $T$ can be represented as follows:
\begin{equation}\label{eq1}
    \Phi=\Delta/T\equiv\Phi(\varphi,\varepsilon),
\end{equation}
where $\Delta$ is the time shift of the disturbed oscillator, $\varphi$ is the
phase difference of the oscillators and $\varepsilon$ is the influence strength.

\begin{figure}[h]
 \includegraphics[width=1\linewidth]{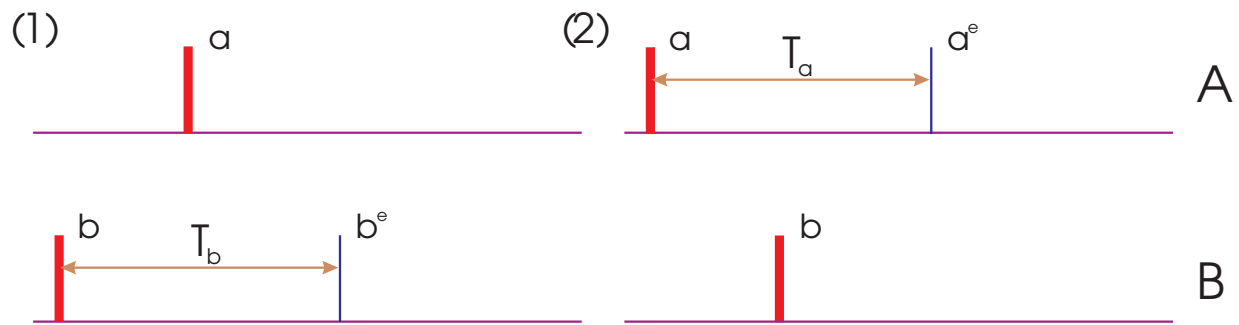}\\
  \caption{ The model of two interacting pulse oscillators.}
  \label{fig1}
\end{figure}

Pacemakers can be represented as a set of separated firing peaks on a time scale.
Assume that the instants of the last firings of the oscillators $A$ and $B$ are
$a$ and $b$ respectively (Fig.~\ref{fig1}). Note that $a$ and $b$ are the moments
of the impacts after all previous phase shifts of the oscillators. In other
words, one can observe the oscillators' firings at these particular instants.
Then it is necessary to analyze two cases:

\begin{enumerate}

\item $b <a$. This is the case 1 in Fig.~\ref{fig1}, i.e. when
the oscillator $B$ has fired before the oscillator $A$. Let us follow the
dynamics of the system in real time. The nearest event, that affects on the
further behavior of the entire system, is an appearance of the pulse $A$ at time
$a$. Let us stop on at this moment and make the forecast. To this end we define a
concept of the moments of \textit{expected} firings of the oscillators, i.e.
$a^e$ and $b^e$. Let us imagine that we have shifted back in time with respect to
the moment $a$. Since the oscillator $A$ has not fired yet, one should expect the
appearance of the next $A$ and $B$ pulses  at the moments $a^e=a$ and $b^e=b+T_b
$ respectively, where $T_b$ is the period of $B$. We call this situation as ``$A$
fires and $B$ is at an expected state'' and denote symbolically as $(a, b^e)$.
Now we consider the instant $a$. Since $A$ fires, the \textit {next} expected
values can be transformed to
\[
\begin{array}{lllll}
a_{next}^{e} & = & a+T_{a} & = & a^{e}+T_{a}, \\
b_{next}^{e} & = & b+T_{b}+\Delta _{b}(\varphi _{b},\varepsilon_b) & = &
b^{e}+\Delta_{b}(\varphi _{b},\varepsilon_b),%
\end{array}%
\]
where $\Delta_{b}(\varphi _{b},\varepsilon_b)$ is the time shift of the
oscillator $B$ due to the impact of $A$. It depends on the phase $\varphi _{b}$
of the pacemaker $A$ with respect to $B$ and the influence strength
$\varepsilon_b$. The phase $\varphi _{b}$ can be calculated as follows
    \[\varphi_b=\frac{a-b}{T_b}\,\, (\textrm{mod}\;1)\]
or, in terms of the expected values,
\[
\varphi_b=\frac{a^e-b^e}{T_b}\,\, (\textrm{mod}\;1).
\]
Note that the phase $\varphi _ {b}$ is a positive value, and it belongs to the
$[0,1]$ segment (negative values in the two previous expressions are eliminated
by the mod 1 operation).

To determine which oscillator fires next, one should compare $a_{next}^e$ and
$b^e_{next}$. If $a_{next}^e<b^e _{next}$, then $A$ fires and $B$ remains at an
expected state until $b^e_{next}$, i.e. the system moves to the state $(a,
b^e)_{next}$. Otherwise, if $b_{next}^e<a^e_{next}$, then $B$ fires, and $A$
jumps to an expected state and the entire system's state becomes $(a^e,
b)_{next}$.

\item $b>a$. This is the case 2 in Fig.~\ref {fig1}. This inverse
situation is analogous to the previous one with the difference in speculations
owing to the firing of the pacemaker $A$ prior to $B$. Then expected values $a^e$
and $b^e$ can be written accordingly: $a^e=a+T_a$ and $b^e=b$. One can call this
case as ``$B $ fires and $A$ is at an expected state'' and denote by $(a^e, b)$.
The next expected values are given by the following expression:
\[
\begin{array}{lllll}
a_{next}^{e} &  = & a^{e}+\Delta _{a}(\varphi _{a},\varepsilon_a), \\
b_{next}^{e} & = & b^e+T_{b},%
\end{array}%
\]
where $\Delta _{a}(\varphi _{a},\varepsilon_a)$ is the time shift of the
oscillator $A$. It depends on the phase of the oscillator $B$ with respect to
$A$, i.e. $\varphi_a=(b^e-a^e)/T_a \,\,(\textrm{mod}\;1)$, and the influence
strength $\varepsilon_a$. Further analysis is also similar to the case 1. Namely,
if $b_{next}^e<a^e _ {next}$, then $B $ fires and $A$ jumps to the expected state
$a^e_{next}$, i.e. the system moves to the state $(a^e, b)_{next}$. Otherwise, if
$a_{next}^e<b^e_{next}$, then  $A$ fires and $B$ remains in the expected state
and the system state becomes $(a, b^e)_{next}$.
\end{enumerate}

Summarizing  the above calculations, the model can be represented by the
following scheme:
\begin{equation}
\begin{array}{*{20}c}
   {\left. {\begin{array}{*{20}c}
   {\left( {\begin{array}{*{20}c}
   a  \hfill \\ {b^e } \hfill  \\
\end{array} } \right)} &  \to  & {\left\{ {\begin{array}{*{20}c}
   {a_{next}^e  = a + T_a }  \\
   {b_{next}^e  = b^e  + \Delta _b (\varphi _b,\varepsilon_b )}  \\
\end{array} } \right.}  \\ \\
   {\left( {\begin{array}{*{20}c}
   {a^e }  \\
   b  \\
\end{array} } \right)} &  \to  & {\left\{ {\begin{array}{*{20}c}
   {a_{next}^e  = a^e  + \Delta _a (\varphi _a,\varepsilon_a )}  \\
   {b_{next}^e  = b + T_b }  \\
\end{array} } \right.}  \\
\end{array} \quad } \right\}} &  \to  & {\left\{ {\begin{array}{*{20}c}
   {\left( {\begin{array}{*{20}c}
   a  \\
   {b^e }  \\
\end{array} } \right)_{next} \;\,\, \textrm{if}\;a_{next}^e  < b_{next}^e }
\\ \\
   {\left( {\begin{array}{*{20}c}
   {a^e }  \\
   b  \\
\end{array} } \right)_{next} \,\, \textrm{if}\;a_{next}^e  > b_{next}^e }  \\
\end{array} } \right.}  \\
\end{array}
\end{equation}

In the notions of the expected values, which we denote for convenience as
$a^e\equiv\hat {a}$ $b^e\equiv\hat {b}$, the dynamics can be described by the
following difference equation:
\begin{equation}
\label{eq2} \left( {{\begin{array}{*{20}c}
 {\widehat {a}_{n + 1}}  \hfill \\
 {\widehat {b}_{n + 1}}  \hfill \\
\end{array}} } \right) = \left( {{\begin{array}{*{20}c}
 {\widehat {a}_{n}}  \hfill \\
 {\widehat {b}_{n}}  \hfill \\
\end{array}} } \right) + \left\{ {{\begin{array}{*{20}c}
 {\left( {{\begin{array}{*{20}c}
 {T_{a}}  \hfill \\
 {\Delta _{b} \left( {\varphi ^{b}_n,\varepsilon_b}  \right)} \hfill \\
\end{array}} } \right),\;\widehat {a}_{n} < \widehat {b}_{n} ,\; {\textrm{and then }}A{\textrm{ fires at time}} \;\widehat{a}_{n}}
\hfill \\ \\
 {\left( {{\begin{array}{*{20}c}
 {\Delta _{a} \left( {\varphi ^{a}_n,\varepsilon_a}  \right)} \hfill \\
 {T_{b}}  \hfill \\
\end{array}} } \right),\;\widehat {b}_{n} < \widehat {a}_{n} ,\; {\textrm{and then }}B{\textrm{ fires at time}} \;\widehat{b}_{n}}, \hfill
\\
\end{array}} } \right.
\end{equation}
where:
$$
\varphi ^{a}_n = \displaystyle\frac{{\left( {\widehat {b}_n - \widehat {a}_n}
\right)}}{{T_{a} }}\;\left( {\textrm{mod}\;1} \right), \ \ \ \ \
 \varphi ^{b}_n = \displaystyle\frac{{\left( {\widehat {a}_n - \widehat {b}_n} \right)}}{{T_{b}
}}\;\left( {\textrm{mod}\;1} \right).
$$

To simulate the dynamics of two interacting oscillators coupled by
PRCs, it is necessary to carry out the iterative
process~(\ref{eq2}) for the expected pulses and put sequentially
in the time scale the firing moments of the oscillators $A$ and
$B$ depending on the result of the comparison $\widehat {a}_n$ and
$\widehat{b}_n$. An investigation of the case of two interacting
pacemakers in more detail with the description of the possible
modes of behavior is given in Section 3.1.

\subsection{Derivation of the basic model equation}

Assume that there are $N$ autonomous pulse oscillators (or pacemakers). Suppose
that all the pacemakers are different. This means that each has its own intrinsic
cycle length $T_i$, $i=1,\ldots,N$, and the beating amplitude. To define coupling
between pacemakers, it is necessary to determine the topology of the system
space. In other words, one should specify the nearest neighbors of each pacemaker
in the space. Vice versa, it is obvious that the determination of coupling
between the elements of such spatially discrete system sets its topology.

First of all we develop the general model of $N$ mutually coupled
pulse oscillators. Suppose that all the pacemakers interact with
each other, i.e. so-called \textit{global coupling} is realized.
The model derived in Section 2.1 can  be easily generalized to the
case of $N$ pacemakers. We operate with the expected values
introduced in Section 2.1, the real firings of the pacemakers are
found by the analysis of the expected impacts series.

\begin{figure}[h]
 \includegraphics[width=0.8\linewidth]{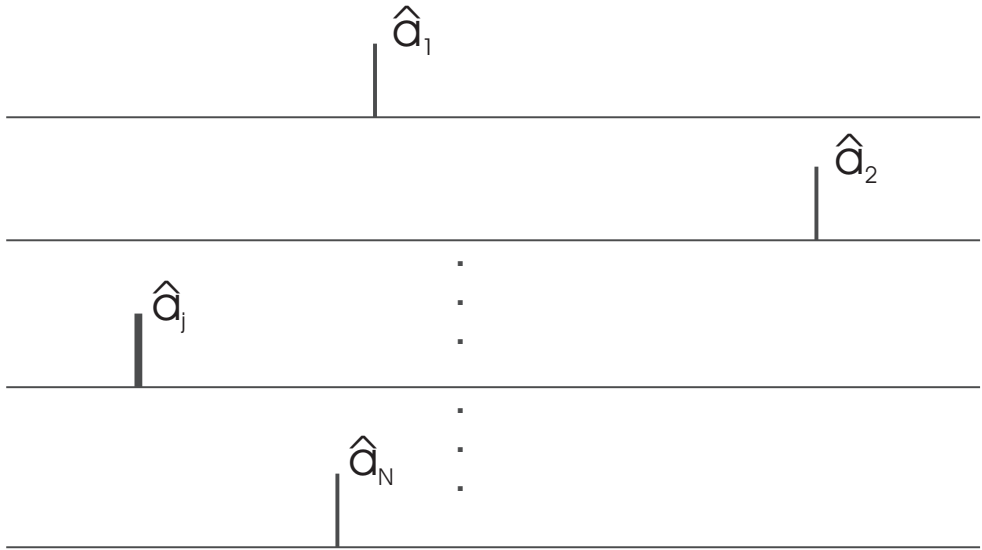}\\
  \caption{ Model of $N$ mutually acting pulse oscillators.}
  \label{fig2}
\end{figure}

Let the set of expected firings $\{{\widehat a}_i \}_{1,\ldots,N}$ be located in
a time axis (see Fig.~\ref{fig2}). This means that in the absence of coupling,
pacemakers strike at these instants. Suppose now that some oscillator acts on
another one by means of the PRC $\{\Delta_
{ij}(\varphi^{ij},\varepsilon_{ij})\}_{1,\ldots,N}$, where $\varphi ^ {ij} $ is
the phase of the $j$-th pacemaker with respect to the $i$-th one and
$\varepsilon_{ij}$ is a total parameter defining coupling between the $j$-th and
$i$-th elements.

The next values of the expected firings can be calculated by the
same manner as in Section 2.1. Because the $j$-th oscillator
appears before all others (see. Fig.~\ref{fig2}), it does not
undergo any influence and fires as a real impact of the $j$-th
pacemaker. Thus, the given event makes the shifts of all other
oscillators according to the set of the PRCs $\{\Delta_{ij}\}$.
The $j$-th oscillator is shifted to a new expected moment as an
unperturbed one, i.e. by adding its own cycle length $T_j$. To get
the next sequential expected values, one should make the same
procedure with the newly obtained expected pulses. The dynamics of
the system can be easily represented as the following iterative
relation:
\begin{equation}\label{eq3}
 \widehat {a}_{n + 1}^{i} = \widehat {a}_{n}^{i} + \left\{
{{\begin{array}{*{20}c}
 {T_{i} ,} \hfill & {i = j,}  \hfill
\\
 {\Delta _{ij} \left( {\varphi _{n}^{ij}},\varepsilon_{ij}  \right),} \hfill & {\;i
\ne j,} \hfill \\
\end{array}} } \right. \;\;j:\;a_{n}^{j} = \min\{ a_{n}^{i}
\}_{i=1,\ldots,N},
\end{equation}
where
\[
\varphi _{n}^{ij} = \displaystyle\frac{{\widehat {a}_{n}^{j} - \widehat
{a}_{n}^{i}} }{{T_{i} }}\;\left( {\textrm{mod}\;1} \right),
\]

It is convenient to rewrite the PRCs by normalizing $\{\Delta_{ij}\}$ on the
intrinsic pacemaker cycle lengths $T_i$ and to define as follows:
\[
\Delta _{ij} \left( {\varphi _{n}^{ij}},\varepsilon_{ij} \right) = \{f_{ij}
\left( {\varphi _{n}^{ij}, \varepsilon _{ij}} \right) \cdot T_{i} \} ,\;\varphi
_{n}^{ij} \in \left[ {0,1} \right],
\]
where $\{f_{ij}(\varphi _{n}^{ij},\varepsilon_{ij})\}$ are the
dimensionless functions, which are also called the PRCs. The real
pacemakers can have identical nature but differ in the intrinsic
cycle lengths. Then the form of dimensionless PRCs is identical
for them, i.e. their $f(\varphi, \varepsilon)$ coincide, while
$\Delta(\varphi,\varepsilon)$ are different. Moreover, it is
convenient to use the functions $\{f _ {ij}(\varphi
_{n}^{ij},\varepsilon_{ij})\}$ in the construction of the
equations for the dimensionless phase differences between
pacemaker pairs. In Section 3 we will demonstrate this approach
for the systems of two and three coupled pulse oscillators.

\section{Applications}

\subsection{Two interacting pulse oscillators}

Let us investigate the system of two interacting pacemakers in
detail. As well as in Section 2.1, suppose that the system
consists of two oscillators $A$ and $B$ coupled by means of the
phase response curves $\Delta_a (\varphi _{n}^{a}, \varepsilon_a)$
and $\Delta_b(\varphi _{n}^{b}, \varepsilon_b)$. It is clear that
the map~(\ref{eq2}) specifying the dynamics of such a system is a
particular case of the general model~(\ref{eq3}) restricted to
$N=2$. Rewrite~(\ref{eq2}) using the expressions for the
dimensionless PRCs:
\begin{equation}\label{eq4}
\left( {{\begin{array}{*{20}c}
 {\widehat {a}_{n + 1}}  \hfill \\
 {\widehat {b}_{n + 1}}  \hfill \\
\end{array}} } \right) = \left( {{\begin{array}{*{20}c}
 {\widehat {a}_{n}}  \hfill \\
 {\widehat {b}_{n}}  \hfill \\
\end{array}} } \right) + \left\{ {{\begin{array}{*{20}c}
 {\left( {{\begin{array}{*{20}c}
 {T_{a}}  \hfill \\
 {f_{b} \left( {\varphi ^{b}_n},\varepsilon_b  \right)}T_b \hfill \\
\end{array}} } \right),\;\widehat {a}_{n} < \widehat {b}_{n} ,\; {\textrm{and then }}A{\textrm{ fires at time}}
\;\widehat{a}_{n}},
\hfill \\ \\
 {\left( {{\begin{array}{*{20}c}
 {f_{a} \left( {\varphi ^{a}_n},\varepsilon_a  \right)}T_a \hfill \\
 {T_{b}}  \hfill \\
\end{array}} } \right),\;\widehat {b}_{n} < \widehat {a}_{n} ,\; {\textrm{and then }}B{\textrm{ fires at time}} \;\widehat{b}_{n}}, \hfill
\\
\end{array}} } \right.
\end{equation}
where:
$$
 \varphi ^{a}_n = \displaystyle\frac{{\left( {\widehat {b}_n - \widehat {a}_n} \right)}}{{T_{a}
}}\;\left( {\textrm{mod}\;1} \right),
 \varphi ^{b}_n = \displaystyle\frac{{\left( {\widehat {a}_n - \widehat {b}_n} \right)}}{{T_{b}
}}\;\left( {\textrm{mod}\;1} \right).
$$

Let us assume that the pacemakers have one the same nature. Hence, one can accept
$f_a(\varphi, \varepsilon)\equiv f_b (\varphi, \varepsilon) \equiv f(\varphi,
\varepsilon)$, where $f(\varphi, \varepsilon)$ is an one-parametric function. The
parameter $\varepsilon$ integrally determines a total influence of one oscillator
on another one. In the symmetric case we have $\varepsilon_a
=\varepsilon_b\equiv\varepsilon$. In our further consideration we do not accept
this symmetry.

From the mathematical point of view the system~(\ref{eq4}) is a map of the real
plane into itself depending on four parameters: $T_a$, $T_b$, $\varepsilon_a$ and
$\varepsilon_b$. It is easy to comprehend that the effective parameter that
changes the behavior of the system~(\ref{eq4}) is the dimensionless ratio
$\delta=T_b/T_a $. Specifying parameters $\varepsilon_a $, $ \varepsilon_b$,
$\delta$ and the function $f(\varphi, \varepsilon)$, one can iterate directly the
equations~(\ref{eq4}) marking on the time axis the values of the real firings of
the pacemakers $A$ and $B$. Some examples of this simulation are presented below.

In spite of the fact that the expression~(\ref{eq4}) defines the two-dimensional
map, the values $\widehat{a}_n$ and $\widehat{b}_n$ increase gradually because
they represent the time series of the expected firings of the oscillators.
Therefore, trajectories of the map~(\ref{eq4}) are infinite. Thus, this is not
informative for us. It is more important to derive a map that determines the
dynamics of the phase difference of the pacemakers.

Let us introduce the dimensionless phase difference of the pacemakers $A$ and
$B$:
\[
x_n =\frac{\widehat {a}_n-\widehat{b} _n}{T_a}.
\]
The choice of $T_a $ as a normalization factor is not essential. Having selected
$T_b$, as a result we obtain the similar expressions. Subtracting the second
equation of the system~(\ref{eq4}) from the first one and dividing the result by
$T_a$ we get the following:
\begin{equation}
\label{eq5}
       x_{n+1}=\left\{
\begin{array}{ll}
    \displaystyle x_n+1-\delta  f\displaystyle\left(\frac{-x_n}{\delta} \,( \textrm{mod}\, 1),\varepsilon_b\right), & x_n<0, \\
    x_n+ f(x_n\, (\textrm{mod}\,1),\varepsilon_a)-\delta, & x_n>0, \\
\end{array}\right.
\end{equation}
where $\delta=T_b/T_a$. Here we take into account that $x_n<0$ if
$\widehat{a}_n<\widehat{b}_n$ and $x_n>0$ if $\widehat{a}_n>\widehat{b}_n$.

\begin{figure}[h]
 \includegraphics[width=0.8\linewidth]{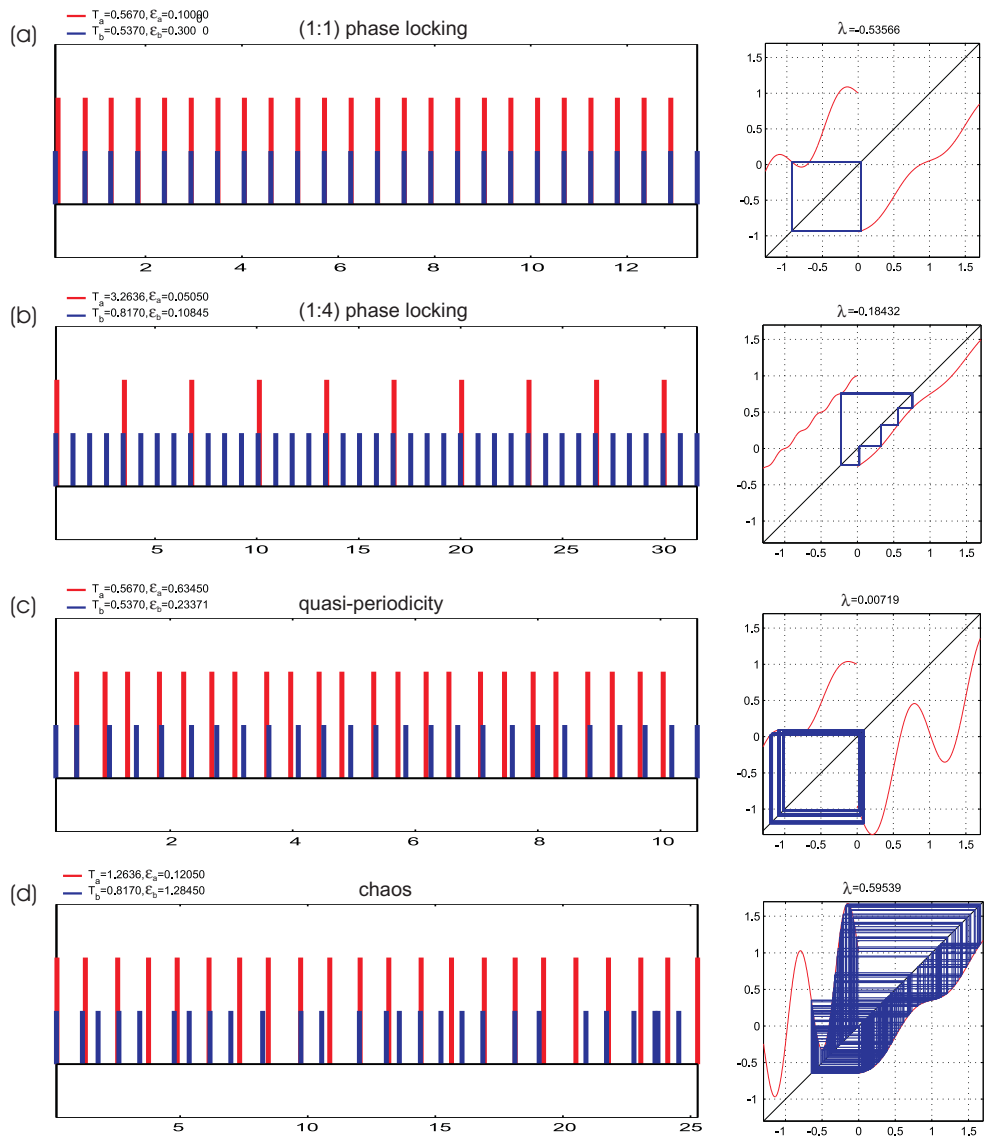}\\
  \caption{ Different types of the behavior of two bidirectionally interacting
pacemakers.}
  \label{fig3}
\end{figure}

Let us make a brief analysis of the developed map. Because in general $x\in
(-\infty; \infty)$, the equation~(\ref{eq5}) represents a one-dimensional
nonlinear map of the real axis into itself. Note that the map can not be reduced
to a circle map by the restriction of $x$ to the range $[0; 1]$ as it is usually
done for two pacemakers interacting by PRCs. It is essentially asymmetric with
respect to changing $x$ to $-x$ (see Fig.~\ref{fig3}). If $f(x,\varepsilon)$ is a
nonmonotonic function, then the map~(\ref{eq5}) is nonlinear, and it can exhibit
a big variety of the behavior: from complex periodic motion to chaotic dynamics.
Owing to $x\in (-\infty; \infty)$, in a rigorous sense it is not a difference of
phases of the pacemakers. In this context one can call $x$  as the
\textit{generalized} phase difference. Analyzing Eq.~(\ref{eq5}) one can find out
which oscillator, $A$ or $B$, fires at the given discrete time $n$. This depends
on the sign of $x$: $A $ fires if $x_n <0$ and  $B $ fires when $x_n> 0$. Thus,
it makes possible to determine the phase-locking degree of the pacemakers.
However, taking into consideration only the values $x_n$, we may not reconstruct
the initial time series of firing events of the pacemakers $A$ and $B$. Using
$x_n$, one can say only about their phase difference.

Let us show how the model equations~(\ref{eq4}),~(\ref{eq5}) can be applied to
the investigation of the behavior of two interacting pacemakers.

Analysis of real systems [Glass {\it et al.}, 1987] shows that the
function $f(x, \varepsilon)$ can take different forms. But, as a
rule, it obeys a number of general properties. For example, $f(0,
\varepsilon)=f(1, \varepsilon)=0$. Usually it has a maximum and a
minimum. Sometimes instead of extrema it has breaks. Let the
function $f(x, \varepsilon)$  be in an elementary form that is
often used (see, e.g. [Glass \& Zeng, 1990]). Namely, let  $f(x,
\varepsilon)=\varepsilon \sin 2\pi x$. This leads to an array of
dynamical equations:
\begin{eqnarray}
\label{eq6}
  \left( {{\begin{array}{*{20}c}
 {\widehat {a}_{n + 1}}  \hfill \\
 {\widehat {b}_{n + 1}}  \hfill \\
\end{array}} } \right) = \left( {{\begin{array}{*{20}c}
 {\widehat {a}_{n}}  \hfill \\
 {\widehat {b}_{n}}  \hfill \\
\end{array}} } \right) + \left\{ {{\begin{array}{*{20}c}
 {\left( {{\begin{array}{*{20}c}
 {T_{a}}  \hfill \\
 {\varepsilon_b\sin \left( 2\pi{\varphi ^{b}_n}  \right)}T_b \hfill \\
\end{array}} } \right),\;\widehat {a}_{n} < \widehat {b}_{n} ,}
\hfill \\
 {\left( {{\begin{array}{*{20}c}
 {\varepsilon_a\sin \left( {2\pi\varphi ^{a}_n}  \right)}T_a \hfill \\
 {T_{b}}  \hfill \\
\end{array}} } \right),\;\widehat {b}_{n} < \widehat {a}_{n} }. \hfill
\\
\end{array}} } \right.  \\ \nonumber \\ \nonumber \\ \label{eq7}
  x_{n+1}=\left\{
\begin{array}{ll}
    \displaystyle x_n+1-\delta  \varepsilon_b \sin\displaystyle(\frac{-x_n}{\delta} \,( \textrm{mod}\, 1)\cdot 2\pi), & x_n<0, \\
    x_n+ \varepsilon_a\sin(x_n\, (\textrm{mod}\,1)\cdot 2\pi)-\delta, & x_n>0. \\
\end{array}\right.
\end{eqnarray}

In Fig.~\ref{fig3} examples of direct simulation of the
system~(\ref{eq6}),~(\ref{eq7}) are presented. In the left column some possible
phase-lockings found on the basis of Eq.~(\ref{eq6}) are displayed. In the right
column the corresponding map~(\ref{eq7}), its periodic orbit and values of
Lyapunov exponents are shown. Fig.~\ref{fig3}d represents the existence of the
chaotic behavior for the system at $\varepsilon_a=0.1205$, $\varepsilon_b=1.2845$
and $\delta=0.6465.$

As a conclusion of this section it should be noted that the system of two
bidirectionally acting pacemakers has been already intensively investigated (see,
e.g., [Ikeda {\it et al.}, 2004] and references herein) as a qualitative model of
cardiac arrhythmia known as modulated parasystole. The interacting oscillators
represented the pairs of cardiac pacemakers such as: the sinoatrial (SA) node and
the ventricle contracting by different factors, e.g. ectopic pacemaker, etc.
Hereby the authors used various kinds of  PRCs $f(x, \varepsilon)$ approximating
the experimental data on stimulating the cardiac cells of animals by single
electric current pulses. Recently in the paper [Loskutov {\it et al.}, 2004] the
system of two interacting pacemakers similar to~(\ref{eq6}),~(\ref{eq7}) has been
analyzed. In this paper taking into account the refractory time the various types
of the smooth PRCs were examined and the bifurcation diagrams of possible
phase-lockings were constructed. However, in [Loskutov {\it et al.}, 2004] the
behavior regimes when the firings of the pacemakers are not alternated, i.e.
cases of strong discrepancies in the intrinsic cycle length ($T_a\gg T_b $ and
vice versa), were not investigated. The system~(\ref{eq6}),~(\ref{eq7}) is more
general and takes into account all possible variants.

\subsection{Three pacemakers}

Below we explain why the special case of three interacting oscillators is worth
individual attention. First, as is known there are three drivers of the rhythm in
the cardiac conductive system: the SA node is a leading pacemaker, the AV
junction and Purkinje fibers are latent pacemakers, which under normal conditions
are suppressed by the sinus rhythm. However, at violations of conductivity and
pulse initiation, the cardiac pacemakers can influence on each other, i.e.
so-called bidirectional coupling may be realized. Second, in pathology a group of
contractile cardiomyocytes can also initiate action potentials: an ectopic
(abnormal) cardiac pacemaker may emerge and start to compete with the SA node for
leading the heart rhythm. Third, at stimulating the cardiac tissue by external
current impulses (cardio-stimulation), the cardiac rate is changed. The external
stimulaters can be naturally included in our general model of interacting
pacemakers as additional leading pulse oscillators.

\begin{figure}[h]
 \includegraphics[width=0.8\linewidth]{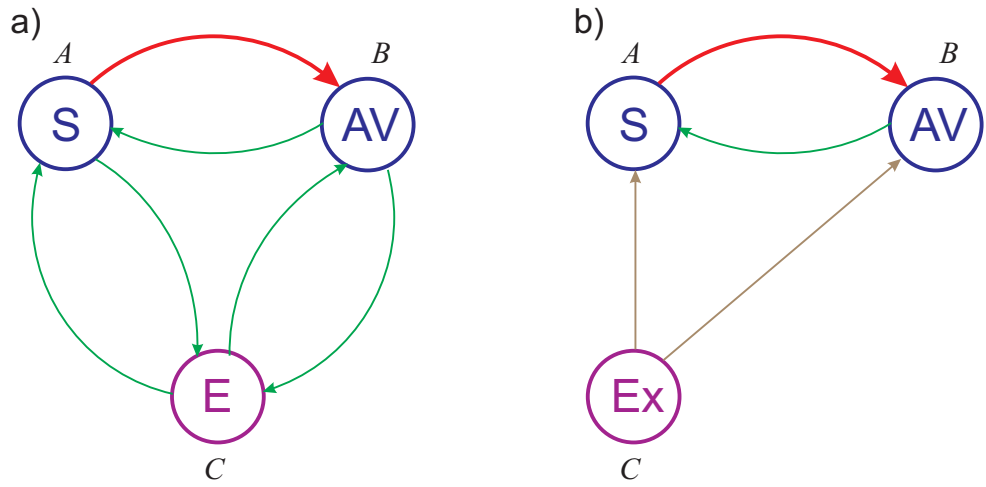}\\
  \caption{ Schematic representation of three pacemakers in the cardiac
tissue.}
  \label{fig4}
\end{figure}

Thus, representation of the heart conductive system as at least
three coupled autonomous oscillators (Fig.~\ref{fig4}) is very
useful for understanding which influence of an additional
pacemaker exerts on a system of two bidirectionally interacting
drivers of the rhythm (i.e. the SA and AV nodes) considered above.
Investigation of the possible behavior modes of a larger amount of
interacting pacemakers turns out to be a very complicated both
analytical (mathematical) and numerical problem. For example for a
system of five coupled pulse oscillators we have 25 various
functions and 29 different parameters (25 values of $\varepsilon _
{ij}$ and 4 independent ratios $T_i/T_j$). This is due to the fact
that, in general, the cardiac pacemakers have various frequencies
(or intrinsic cycle lengths $T_i$ ) and the different nature. This
means that the PRCs $\{f_ {ij}(x,\varepsilon_{ij})\}$ determining
coupling between a pair of pacemakers have different forms. For
some heart pacemakers PRCs have been measured by the direct
experiments on cardiomyocytes of animals (see [Glass {\it et al.},
1986]). Other PRCs can be chosen using general principles based on
the nature of nodes or on the basis of the collateral measurements
[Ikeda {\it et al.}, 1988].

Applying the general model equations~(\ref{eq3}) for three interacting
pacemakers, one can get the following system:
\begin{equation}\label{eq8}
\left( {{\begin{array}{*{20}c}
 {\widehat{a}_{n + 1}}  \hfill \\
 {\widehat{b}_{n + 1}}  \hfill \\
 {\widehat{c}_{n + 1}}  \hfill \\
\end{array}} } \right) = \left( {{\begin{array}{*{20}c}
 {\widehat{a}_{n}}  \hfill \\
 {\widehat{b}_{n}}  \hfill \\
 {\widehat{c}_{n}}  \hfill \\
\end{array}} } \right) + \left\{ {{\begin{array}{*{20}c}
 {\left( {{\begin{array}{*{20}c}
 {T_{a}}  \hfill \\
 {\Delta _{ba} \left( {\varphi _{n}^{ba}},\varepsilon_{ba}  \right)} \hfill \\
 {\Delta _{ca} \left( {\varphi _{n}^{ca}},\varepsilon_{ca}  \right)} \hfill \\
\end{array}} } \right),\,\textrm{if}\,\,\widehat{a}_{n} < \widehat{b}_{n}
,\widehat{c}_{n} ;\quad A\;\textrm{fires at time}\;\widehat{a}_{n}}  \hfill \\
 {} \hfill \\
 {\left( {{\begin{array}{*{20}c}
 {\Delta _{ab} \left( {\varphi _{n}^{ab}},\varepsilon_{ab}  \right)} \hfill \\
 {T_{b}}  \hfill \\
 {\Delta _{cb} \left( {\varphi _{n}^{cb}},\varepsilon_{cb}  \right)} \hfill \\
\end{array}} } \right),\,\textrm{if}\,\,\widehat{b}_{n} < \widehat{a}_{n}
,\widehat{c}_{n} ;\quad B\;\textrm{fires at time}\;\widehat{b}_{n}}  \hfill \\
 {} \hfill \\
 {\left( {{\begin{array}{*{20}c}
 {\Delta _{ac} \left( {\varphi _{n}^{ac}},\varepsilon_{ac}  \right)} \hfill \\
 {\Delta _{bc} \left( {\varphi _{n}^{bc}},\varepsilon_{bc}  \right)} \hfill \\
 {T_{c}}  \hfill \\
\end{array}} } \right),\,\textrm{if}\,\,\widehat{c}_{n} < \widehat{a}_{n}
,\widehat{b}_{n} ;\quad C\;\textrm{fires at time}\;\widehat{c}_{n}}
\end{array}} } \right.
\end{equation}
where
\[\begin{array}{l}
 \varphi _{n}^{ba} = \displaystyle\frac{{\widehat{a}_{n} - \widehat{b}_{n}} }{{T_{b}
}}\;\left( {\textrm{mod}\;1} \right),\quad \varphi _{n}^{ca} =
\displaystyle\frac{{\widehat{a}_{n} -
\widehat{c}_{n}} }{{T_{c}} }\;\left( {\textrm{mod}\;1} \right) \\
 \varphi _{n}^{ab} = \displaystyle\frac{{\widehat{b}_{n} - \widehat{a}_{n}} }{{T_{a}
}}\;\left( {\textrm{mod}\;1} \right),\quad \varphi _{n}^{cb} =
\displaystyle\frac{{\widehat{b}_{n} -
\widehat{c}_{n}} }{{T_{c}} }\;\left( {\textrm{mod}\;1} \right) \\
 \varphi _{n}^{ac} = \displaystyle\frac{{\widehat{c}_{n} - \widehat{a}_{n}} }{{T_{a}
}}\;\left( {\textrm{mod}\;1} \right),\quad \varphi _{n}^{bc} =
\displaystyle\frac{{\widehat{c}_{n} -
\widehat{b}_{n}} }{{T_{b}} }\;\left( {\textrm{mod}\;1} \right) \\
 \end{array}
 \]
The response functions $\Delta_{ij}$ are supposed to have a form
$\Delta_{ij}(\varphi _{n}^{ij},\varepsilon_{ij})=f_{ij}(\varphi
_{n}^{ij},\varepsilon_{ij})T_i,\, \varphi _{n}^{ij}\in[0;1], \, i,j=a,b,c$. The
example of a system of three bidirectionally interacting cardiac pacemakers: the
SA node, the AV junction and ectopic pacemaker, is shown in Fig.~\ref{fig4}a.

\begin{figure}[h]
 \includegraphics[width=0.8\linewidth]{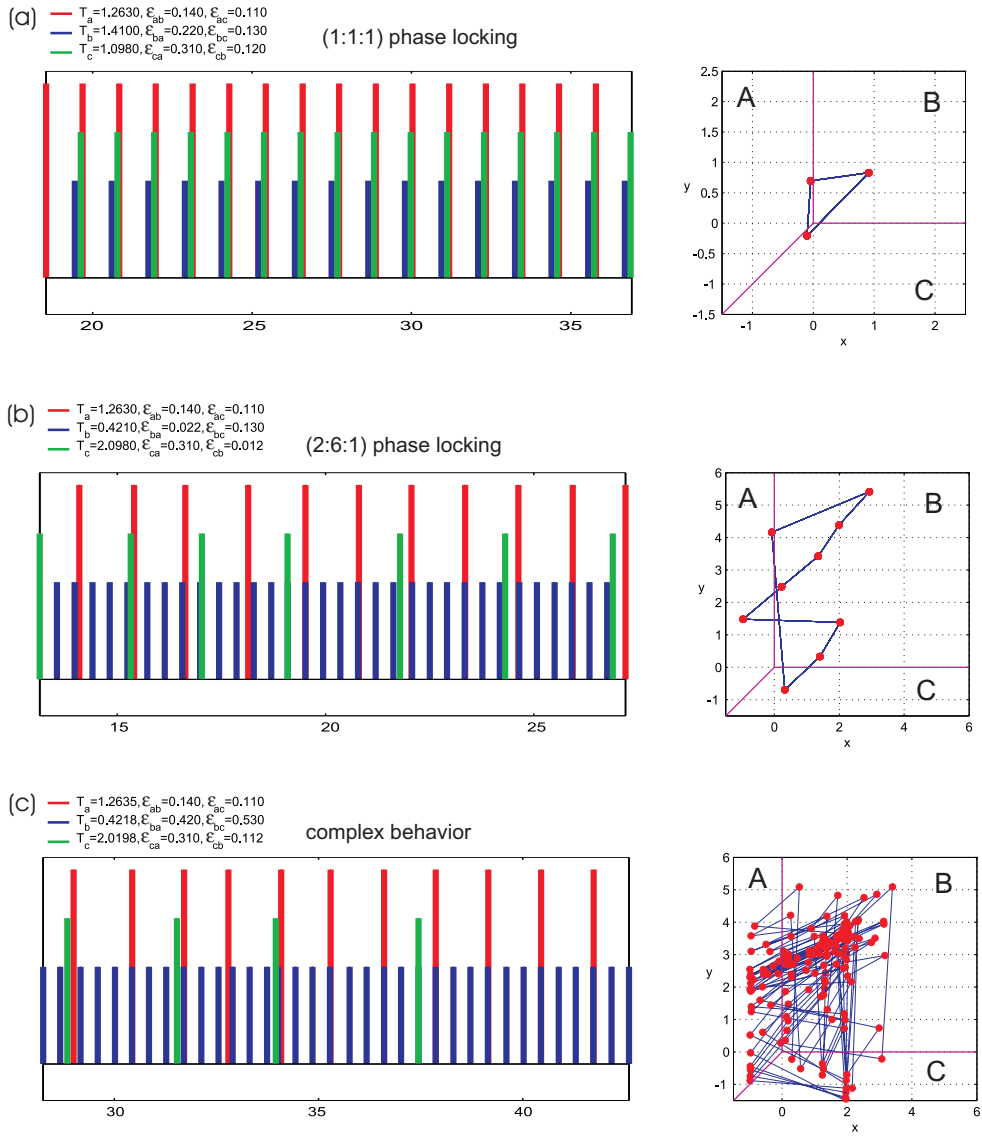}\\
  \caption{ Different types of the behavior of three pulse oscillators.}
  \label{fig5}
\end{figure}

It is convenient to study Eqs.~(\ref{eq8}) introducing phase differences of the
pacemakers. Let us define,
\[
 x_{n} = \displaystyle\frac{{\widehat{a}_{n} - \widehat{b}_{n}} }{{T_{b}} },\ \ \ \ \
  y_{n} =
\displaystyle\frac{{\widehat{c}_{n} - \widehat{b}_{n}} }{{T_{b}} },\ \ \ \ \
 \alpha = \displaystyle\frac{{T_{a}} }{{T_{b}} }, \ \ \ \ \
\beta =\displaystyle\frac{{T_{c}} }{{T_{b}} }.
\]
Then for the phase differences we obtain the map of the real plane into itself:
\begin{equation}\label{eq9}
\begin{array}{*{20}l}\left( {{\begin{array}{*{20}c}
 {x_{n + 1}}  \hfill \\
 {y_{n + 1}}  \hfill \\
\end{array}} } \right) =  \left( {{\begin{array}{*{20}c}
 {x_{n}}  \hfill \\
 {y_{n}}  \hfill \\
\end{array}} } \right) + \\ \\  +\left\{ {{\begin{array}{*{20}c}
 {\left( {{\begin{array}{*{20}c}
 {\alpha -  f_{ba} \left( {\displaystyle\left\{x_{n} \right\},
 \varepsilon _{ba}}  \right)} \hfill \\
 {\beta f_{ca} \left( {\displaystyle\left\{\frac{{ x_{n}-y_n} }{{\beta} }\right\},\varepsilon _{ca}}  \right) -  f_{ba} \left(
\left\{x_{n}\right\},\varepsilon _{ba}  \right)} \hfill \\
\end{array}} } \right),\quad x_{n} < 0,\;x_{n}   < y_{n}} \hfill
\\ \\
 {\left( {{\begin{array}{*{20}c}
 \alpha{f_{ab} \left( \left\{\displaystyle\frac{-x_{n}}{\alpha}\right\} ,\varepsilon _{ab}  \right)
 -1
}  \hfill \\
 {\beta f_{cb} \left( {\displaystyle\left\{\frac{{ - y_{n}} }{{\beta} }\right\},\varepsilon _{cb}}  \right)}-1 \hfill \\
\end{array}} } \right),\quad x_{n} > 0,\;y_{n} > 0} \hfill \\ \\
 {\left( {{\begin{array}{*{20}c}
 \alpha{f_{ac} \left( {\left\{ \displaystyle\frac{y_{n}-x_{n}}{\alpha}  \right\},\varepsilon _{ac}}  \right) - f_{bc} \left( \left\{y_{n}
\right\},\varepsilon _{bc}  \right)} \hfill \\
\beta- {f_{bc} \left( \left\{y_{n} \right\},\varepsilon _{bc} \right)
}  \hfill \\
\end{array}} } \right),\quad  y_{n} < x_{n} ,\;y_{n} < 0} \hfill \\
\end{array}} } \right.\end{array}
\end{equation}
Here we denote $\{x\}$ as the fractional part of $x$. The conditions in the right
hand side of~(\ref{eq9}) divide the plane into three areas corresponding to the
firing pacemaker (see Fig.~\ref{fig5}). The map~(\ref{eq9}) has breaks on the
boundaries of these areas.

Suppose that the pacemaker $B$ is an external stimulus (see Fig.~\ref{fig4}b).
Then the expressions~(\ref{eq8}),~(\ref{eq9}) become simpler. The stimulus $B$
acts on the pacemakers $A$ and $C$ but does not experience any influence. In this
situation $f_{ba}(x,\varepsilon_{ba})\equiv f_{bc}(x,\varepsilon_{bc})\equiv 0$
and the model~(\ref{eq8}),~(\ref{eq9}) can be rewritten as follows:
\begin{equation}\label{eq10}
\left( {{\begin{array}{*{20}c}
 {\widehat{a}_{n + 1}}  \hfill \\
 {\widehat{b}_{n + 1}}  \hfill \\
 {\widehat{c}_{n + 1}}  \hfill \\
\end{array}} } \right) = \left( {{\begin{array}{*{20}c}
 {\widehat{a}_{n}}  \hfill \\
 {\widehat{b}_{n}}  \hfill \\
 {\widehat{c}_{n}}  \hfill \\
\end{array}} } \right) + \left\{ {{\begin{array}{*{20}c}
 {\left( {{\begin{array}{*{20}c}
 {T_{a}}  \hfill \\
 {0} \hfill \\
 {\Delta _{ca} \left( {\varphi _{n}^{ca}}, \varepsilon _{ca}  \right)} \hfill \\
\end{array}} } \right),\,\textrm{if}\,\,\widehat{a}_{n} < \widehat{b}_{n}
,\widehat{c}_{n} ;\quad A\;\textrm{fires at time}\;\widehat{a}_{n}}  \hfill \\
 {} \hfill \\
 {\left( {{\begin{array}{*{20}c}
 {\Delta _{ab} \left( {\varphi _{n}^{ab}}, \varepsilon _{ab}  \right)} \hfill \\
 {T_{b}}  \hfill \\
 {\Delta _{cb} \left( {\varphi _{n}^{cb}}, \varepsilon _{cb}  \right)} \hfill \\
\end{array}} } \right),\,\textrm{if}\,\,\widehat{b}_{n} < \widehat{a}_{n}
,\widehat{c}_{n} ;\quad B\;\textrm{fires at time}\;\widehat{b}_{n}}  \hfill \\
 {} \hfill \\
 {\left( {{\begin{array}{*{20}c}
 {\Delta _{ac} \left( {\varphi _{n}^{ac}}, \varepsilon _{ac}  \right)} \hfill \\
 {0} \hfill \\
 {T_{c}}  \hfill \\
\end{array}} } \right),\,\textrm{if}\,\,\widehat{c}_{n} < \widehat{a}_{n}
,\widehat{b}_{n} ;\quad C\;\textrm{fires at time}\;\widehat{c}_{n}}
\end{array}} } \right.
\end{equation}

The corresponding expression for the phase differences  takes the form:
\begin{equation}\label{eq11}
\begin{array}{*{20}l}\left( {{\begin{array}{*{20}c}
 {x_{n + 1}}  \hfill \\
 {y_{n + 1}}  \hfill \\
\end{array}} } \right) =  \left( {{\begin{array}{*{20}c}
 {x_{n}}  \hfill \\
 {y_{n}}  \hfill \\
\end{array}} } \right) + \\ \\  +\left\{ {{\begin{array}{*{20}c}
 {\left( {{\begin{array}{*{20}c}
 {\alpha } \hfill \\
 {\beta f_{ca} \left( {\displaystyle\left\{\frac{{ x_{n}-y_n} }{{\beta} }\right\},\varepsilon _{ca}}  \right) } \hfill \\
\end{array}} } \right),\quad x_{n} < 0,\;x_{n}   < y_{n}} \hfill
\\ \\
 {\left( {{\begin{array}{*{20}c}
 \alpha{f_{ab} \left( \left\{\displaystyle\frac{-x_{n}}{\alpha}\right\} ,\varepsilon _{ab}  \right)
 -1
}  \hfill \\
 {\beta f_{cb} \left( {\displaystyle\left\{\frac{{ - y_{n}} }{{\beta} }\right\},\varepsilon _{cb}}  \right)}-1 \hfill \\
\end{array}} } \right),\quad x_{n} > 0,\;y_{n} > 0} \hfill \\ \\
 {\left( {{\begin{array}{*{20}c}
 \alpha{f_{ac} \left( {\left\{ \displaystyle\frac{y_{n}-x_{n}}{\alpha}  \right\},\varepsilon _{ac}}  \right) } \hfill \\
\beta  \hfill \\
\end{array}} } \right),\quad  y_{n} < x_{n} ,\;y_{n} < 0.} \hfill \\
\end{array}} } \right.\end{array}
\end{equation}

In Fig.~\ref{fig5} some examples of phase-lockings and complex dynamics for the
system ~(\ref{eq8}),~(\ref{eq9}) are shown. The PRCs are chosen in the sinus
form, i.e. $f_{ij}(x,\varepsilon_{ij})=\varepsilon_{ij}\sin(2\pi x)$. $T_i$ and
$\varepsilon_{ij}$ are indicated in the legends. It is obvious that the model of
three interacting pacemakers requires further investigation in detail. In
particular, we believe that appropriate choice of influence strength and period
of external stimulation in the equations~(\ref{eq11}) will remove the system from
undesirable complex behavior to the normal heart rhythm. It will be represented
within forthcoming works.

\section{Lattices of Coupled Pulse Oscillators}

Let us demonstrate briefly the way how one can approximate discrete distributed
media on the basis of the general model of coupled oscillators~(\ref {eq3}).
Considering a cardiac pacemaker on the microscopic level, one can represent it as
a large group of cells, which generate the heart rhythm and synchronize their
action potentials to initiate the cardiac contraction. Thus, instead of examining
a single pacemaker we should construct a lattice of coupled pulse oscillators. In
this paper we restrict ourselves to the one- (chain) and two-dimensional
(lattice) cases.

Let us suppose that the autonomous pacemakers are located in sites of the
two-dimensional square lattice of the size $N\times M$. An element of the lattice
with coordinates $(i, j)$ is designated as $A^{i, j}$, where $i=1,\ldots,N$ and
$j=1,\ldots,M$. We restrict consideration to the homogeneous medium and accept a
number of limitations on anisotropy. This means that the pacemakers of the
lattice are identical, i.e. they have the same cycle length $T_{i, j}\equiv T$,
$i=1,\ldots,N,\,j=1,\ldots,M$ (however, in fact, cells from the periphery of the
sinus pacemaker show the shortest cycle length, although the centre acts as the
leading pacemaker site). This limitation decreases a quantity of parameters of
the system and hence simplifies the study of the model. Now we should define
coupling between elements. In works devoted to the coupled map lattices (CML) two
main types of coupling are usually assumed: the nearest neighbors and global
couplings (see, e.g., [Kaneko, 1989]). Since in the previous sections we supposed
that pacemakers interact with each other, this time as an example we consider
lattices with the coupling to nearest neighbors: first, a two-dimensional
lattice, and then a chain of coupled pulse oscillators.

\begin{figure}[h]
 \includegraphics[width=0.8\linewidth]{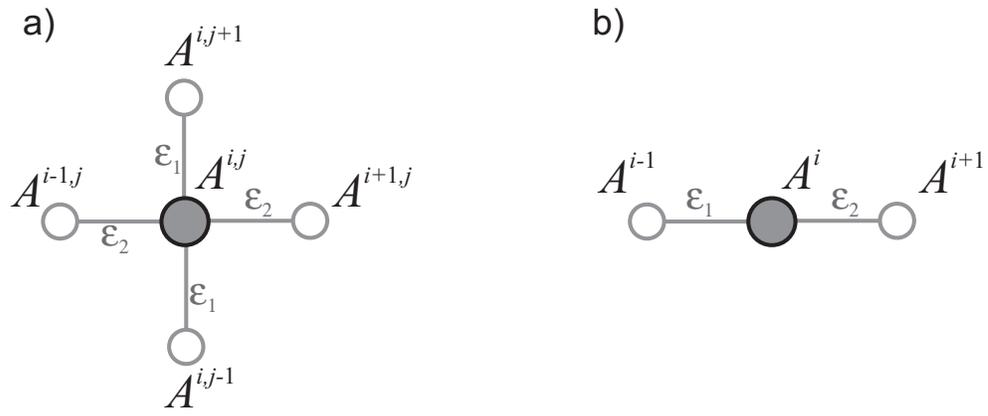}\\
  \caption{ $2D$ and $1D$ lattices of coupled pulse oscillators.}
  \label{fig6}
\end{figure}

Within the square lattice each pacemaker $A^{i, j}$ interacts with four of the
neighbors according to the schematic picture in Fig.~\ref{fig6}a. Taking into
account the limitation of homogeneous medium, one can assert that all couplings
of the lattice are identical, i.e. each two adjacent elements interact with each
other by a general law defined by the identical PRCs $f(x,\varepsilon)$.
Moreover, suppose that the coupling between a pair of elements is isotropic in
such a sense that $\varepsilon_{(i, j)(i', j')}=\varepsilon_{(i ', j ')(i, j)}$,
and is equal to one of the values $\varepsilon_1$ or $\varepsilon_2 $ depending
on the relative position of elements. This means that there is an anisotropy of
the influence strength in vertical and horizontal directions. In other words, if
two pacemakers are the neighbors in the vertical direction, they interact by
$f(x, \varepsilon_1)$, and if they are horizontal neighbors, then they are
coupled by $f(x,\varepsilon_2)$.

Note that all these restrictions are made for the simplification of the
analytical form of the resulting model. In general, it is possible to write the
expressions for a two-dimensional lattice of coupled pulse oscillators without
any limitations. It is a subject of a standalone investigation and lies beyond
the framework of the given study.

Let us write equations that determine the iterative dynamics of
the expected firings of the pacemakers
$\{\widehat{a}^{i,j}\}_{^{i=1,\ldots,N}_{j=1,\ldots,M}}$ on the
basis of the approach represented in Sec.2.2. To obtain the
$(n+1)$-th value of the individual element $\widehat{a}^{i, j}$,
it is necessary to analyze all elements of the lattice since they
are coupled with each other by means of a local coupling. In other
words, the considered element cannot be affected by others at the
$n$-th time step because the latter is suppressed by the influence
of other elements  and remains at the expected state until the
$(n+1)$-th step. Then dynamics of the model can be described by
the following expression:
\begin{equation}\label{eq12}
\begin{array}{ll}
\widehat{a}_{n+1}^{i,j}=\widehat{a}_{n}^{i,j}+ & \left\{
\begin{array}{ll}
T, & {\textrm{if}}~\widehat{a}_{n}^{i,j}=\widehat{a}^{\min }, \\
f(\varphi _{n}^{(i,j)(0,+1)},\varepsilon _{1})\cdot T, & {\textrm{if}}~\widehat{a%
}_{n}^{i,j+1}=\widehat{a}^{\min }, \\
f(\varphi _{n}^{(i,j)(0,-1)},\varepsilon _{1})\cdot T, & {\textrm{if}}~\widehat{a%
}_{n}^{i,j-1}=\widehat{a}^{\min }, \\
f(\varphi _{n}^{(i,j)(+1,0)},\varepsilon _{2})\cdot T, & {\textrm{if}}~\widehat{a%
}_{n}^{i+1,j}=\widehat{a}^{\min }, \\
f(\varphi _{n}^{(i,j)(-1,0)},\varepsilon _{2})\cdot T, & {\textrm{if}}~\widehat{a%
}_{n}^{i-1,j}=\widehat{a}^{\min }, \\
0, & {\textrm{otherwise}}%
\end{array}%
\right. \widehat{a}^{\min }=\min \{\widehat{a}_{n}^{i,j}\}_{\
_{j=1...M}^{i=1...N}}%
\end{array}%
\end{equation}
where the phases are
\[
\begin{array}{l}
\varphi _{n}^{(i,j)(0,+1)}=\left\{ \displaystyle\frac{\widehat{a}_{n}^{i,j+1}-\widehat{a}%
_{n}^{i,j}}{T}\right\}  \\ \\
\varphi _{n}^{(i,j)(0,-1)}=\left\{ \displaystyle\frac{\widehat{a}_{n}^{i,j-1}-\widehat{a}%
_{n}^{i,j}}{T}\right\}  \\ \\
\varphi _{n}^{(i,j)(+1,0)}=\left\{ \displaystyle\frac{\widehat{a}_{n}^{i+1,j}-\widehat{a}%
_{n}^{i,j}}{T}\right\}  \\ \\
\varphi _{n}^{(i,j)(-1,0)}=\left\{ \displaystyle\frac{\widehat{a}_{n}^{i-1,j}-\widehat{a}%
_{n}^{i,j}}{T}\right\}
\end{array}%
\]

The constructed model demands detailed investigation on the basis of the approach
developed for the CML (see, e.g., [Kaneko, 1989]). It will be represented in the
succeeding works.

As the second example let us consider a chain of the identical pulse oscillators
coupled by the nearest neighbor principle. We restrict ourselves to a homogeneous
case with anisotropy of the right and left direction in the influence strength
between the nearest neighbors. The schematic picture of the chain is shown in
Fig.~\ref{fig6}b. Similarly to the above consideration one can get:
\begin{equation}\label{eq13}
\begin{array}{ll}
\widehat{a}_{n+1}^{i}=\widehat{a}_{n}^{i}+ & \left\{
\begin{array}{ll}
T, & {\textrm{if}}~\widehat{a}_{n}^{i}=\widehat{a}^{\min }, \\
f(\varphi _{n}^{i,+1},\varepsilon _{2})\cdot T, & {\textrm{if}}~\widehat{a}%
_{n}^{i+1}=\widehat{a}^{\min }, \\
f(\varphi _{n}^{i,-1},\varepsilon _{1})\cdot T, & {\textrm{if}}~\widehat{a}%
_{n}^{i-1}=\widehat{a}^{\min }, \\
0, & {\textrm{otherwise}}%
\end{array}%
\right. \widehat{a}^{\min }=\min \{\widehat{a}_{n}^{i}\}_{i=1...N}%
\end{array}%
\end{equation}
where the phases are the following
\[
\begin{array}{l}
\varphi _{n}^{i,+1}=\left\{ \displaystyle\frac{\widehat{a}_{n}^{i+1}-\widehat{a}_{n}^{i}}{%
T}\right\}  \ \ \ \
\varphi _{n}^{i,-1}=\left\{ \displaystyle\frac{\widehat{a}_{n}^{i-1}-\widehat{a}_{n}^{i}}{%
T}\right\}.
\end{array}%
\]
If $\varepsilon_1=0$ (or $\varepsilon_2=0$), then Eqs~(\ref{eq13}) define the
so-called \textit{open-flow} model [Willeboordse \& Kaneko, 1994].

Because in this work we present a general approach of developing models without
the detailed analysis of their behavior, the type of boundary conditions for both
lattices has not been indicated. Hence, to investigate such systems analytically
or numerically, one should set the boundary conditions along with the PRCs $f(x,
\varepsilon) $. Usually the boundary conditions are chosen as periodic, i.e.
$\widehat{a}^{i, j+M}\equiv \widehat{a}^{i, j};\,\widehat{a}^{i+N, j}\equiv
\widehat {a} ^ {i, j} $ for the two-dimensional lattice and $\widehat
{a}^{i+N}\equiv \widehat {a} ^i $ for the one-dimensional one. For the open-flow
model a condition of the fixed left boundary, $\widehat{a}^1_n\equiv const$, is
frequently accepted.

The described models~(\ref{eq3}), (\ref{eq12}) and~(\ref{eq13}) admit
generalization to a natural inhomogeneous case by placing different intrinsic
cycle lengths of the pacemakers, PRCs and influence strengths for various groups
of elements. However, consideration of inhomogeneous anisotropic lattices is
extremely difficult problem even for numerical analysis. The first attempts of
investigating inhomogeneous lattices of coupled maps (ICML) were described in
[Vasil'ev {\it et al.}, 2000; Loskutov {\it et al.}, 2002; Rybalko \& Loskutov,
2004].

\section{Summary and Limitations}

In the present study we propose a quite general discrete model of
active media by introducing a simple phase response curve
interaction between leading centers. We have shown that the PRC
can be a useful ``tool'' for representation of the interaction
between pacemakers in cardiac tissue both on a large and small
scales. This PRC based model together with demonstrating complex
(chaotic) behavior, can describe the entrainment and
synchronization phenomena of interacting pulse oscillators. It can
also aid to understand their response to an external stimulus with
variable intensity and duration (see Fig.~\ref{fig4}b), as
previously observed in experimental studies [Jalife {\it et al.},
1976; Jalife {\it et al.}, 1980].

Starting with consideration of two interacting pulse oscillators and introducing
new concepts of expected values, we have extrapolated our PRC based approach to
investigate the mutual influence among an arbitrary large ensemble of pacemakers.
The specific cases of the proposed model show that it can be very useful for
investigating the dynamical interaction of cardiac nodes.

The last part of our study suggests that the derived general model can be easily
applied to construct one-- and two--dimensional lattices of active elements
interacting by the nearest neighbors type. Extension of the model to a
three--dimensional case is straightforward.

Finally, some limitations of our approach should be mentioned. First, the
proposed model is not complete, there is no a time delay in pulse propagation
among pacemakers, which can be very important for describing cardiac arrhythmias.
Second, we represented cardiac tissue as a discrete one and used iterative
approach to investigate its behavior. However, a large amount of realistic
examples of active media is treated as continuous. Nevertheless, cardiac tissue
is not a continuum, but is built up by discrete cardiomyocytes (or nodes with
approximate dimensions 0.15 mm $\times$ 0.02 mm $\times$ 0.01 mm) [Kuramoto,
1984].

Third and most important, to analyze the essential features governing dynamics of
network of active elements, we have not included many important properties of the
real conductive cells. These include the relaxation after stimulating, the
prolonged (non-peak) form of pulses profiles, realistic topological structure,
etc. Further investigations are required to incorporate these features to the
general combined model.

\section{Acknowledgements}

This paper was partially supported by INTAS fellowship No 03-55-1920, granted to
Ekaterina Zhuchkova. Also, we would like to thank Prof.Alexander Loskutov for
critical reading of the manuscript.

\newpage

{\Large {\bf References}}

\bigskip

\parindent=-15pt

Antzelevitch, C., Jalife, J. \& Moe, G. K. [1982] {``Electrotonic modulation of
pacemaker activity - further biological and mathematical observations on the
behavior of modulated parasystole,''} {\em Circulation} {\bf 66}(6), 1225--1232.

Bub, G. \& Glass, L. [1994] {``Bifurcations in a continuous circle map: A theory
for chaotic cardiac arrhythmia,''} {\em Int. J. Bifurcation and Chaos} {\bf
5}(2), 359--371.

Clayton, R. H. [2001] {``Computational models of normal and abnormal action
potential propagation in cardiac tissue: linking experimental and clinical
cardiology,''} {\em Physiol. Meas.} {\bf 22}, R15--R34.

Clayton, R. H., Zhuchkova, E. A. \& Panfilov, A. V. [2006] {``Phase singularities
and filaments: Simplifying complexity in computational models of ventricular
fibrillation,''} {\em Prog. Biophys. Mol. Biol.} {\bf 90}, 378--398.

Courtemanche,  M.,  Glass, L., Belair, J.,  Scagliotti, D. \& Gordon, D. [1989]
{``A circle map in a human heart,''} {\em Physica D} {\bf 49}, 299--310.

Glass, L., Goldberger, A. L. \& Belair, J. [1986] {``Dynamics of pure
parasystole,''} {\em Am. J. Physiol.} {\bf 251}(4), H841--H847.

Glass, L., Goldberger, A. L., Courtemanche, M. \& Shrier, A. [1987] {``Nonlinear
dynamics, chaos and complex cardiac arrhythmias,''} {\em Proc. R. Soc. London
Ser. A-Math. Phys. Eng. Sci.} {\bf 413}(1844), 9--26.

Glass, L. \& Zeng, W. Z. [1990] {``Complex bifurcations and chaos in simple
theoretical models of cardiac oscillations,''} {\em Ann. N.Y. Acad. Sci.} {\bf
591}, 316--327.

Glass, L., Nagai, Yo.,  Hall, K., Talajie, M. \& Nattel, S. [2002] {``Predicting
the entrainment of reentrant cardiac waves using phase resetting curves,''} {\em
Phys. Rev. E} {\bf 65}, 021908-1--021908-10.

Goldberger, A. L. [1990] {``Nonlinear dynamics, fractals and chaos: Applications
to cardiac electrophysiology,''} {\em Ann. Biomed. Eng.} {\bf 18}(2), 195--198.

Guevara, M. R. \& Shrier, A. [1987] {``Phase resetting in a model of cardiac
Purkinje fiber,''} {\em Biophys. J.} {\bf 52}(2), 165--175.

Ikeda, N. [1982] {``Model of bidirectional interaction between myocardial
pacemakers based on the phase response curve,''} {\em Biol. Cybern.} {\bf 43}(3),
157--167.

Ikeda, N., DeLand, E., Miyahara, H., Takeuchi, A., Yamamoto, H. \& and Sato, T.
[1988] {``A personal computer-based arrhythmia generator based on mathematical
models of cardiac arrhythmia,''} {\em J. Electrocardiol.} {\bf 21}(Suppl).

Ikeda, N., Takeuchi, A., Hamada, A., Goto, H., Mamorita, N. \& Takayanagi, K.
[2004] {``Model of bidirectional modulated parasystole as a mechanism for cyclic
bursts of ventricular premature contractions,''} {\em Biol. Cybern.} {\bf 91}(1),
37--47.

Jalife, J. \& Moe, G. K. [1976] {``Effects of electronic potentials on pacemaker
activity of canine Purkinje fibers in relation to parasystole,''} {\em Circ.
Res.} {\bf 39}(6), 801--808.

Jalife, J., Hamilton, A. J., Lamanna, V. R. \& and Moe, G. K. [1980] {``Effects
of current flow on pacemaker activity of the isolated kitten sinoatrial node,''}
{\em Am. J. Physiol.} {\bf 238}(3), H307--H316.

Kaneko, K. [1989] {``Spatiotemporal chaos in one-dimensional and two-dimensional
coupled map lattices,''} {\em Physica D} {\bf 37}(1-3), 60--82.

Kaneko, K. [1990] {``Clustering, coding, switching, hierarchical ordering, and
control in a network of chaotic element,''} {\em Physica D} {\bf 41}(2),
137--172.

Kuramoto, Y. [1984] {\em Chemical Oscillations, Waves, and Turbulence}
(Springer-Verlag, Berlin).

Kuramoto, Y. [1995] {``Scaling behavior of turbulent oscillators with nonlocal
interaction,''} {\em Prog. Theor. Phys.} {\bf 94}(3), 321--330.

Loskutov, A., Prokhorov, A. K. \& Rybalko, S. D. [2002] {``Dynamics of
inhomogeneous chains of coupled quadratic maps,''} {\em Theor. Math. Phys.} {\bf
132}(1), 983--999.

Loskutov, A., Rybalko, S. \& Zhuchkova, E. [2004] {``Model of cardiac tissue as a
conductive system with interacting pacemakers and refractory time,''} {\em Int.
J. Bifurcation and Chaos} {\bf 14}(7), 2457--2466.

Reiner, V. S. \& Antzelevich, C. [1985] {``Phase resetting and annihilation in a
mathematical model of sinus node,''} {\em Am. J. Physiol.} {\bf 249},
H1143--H1153.

Rybalko, S. \& Loskutov, A. [2004] {``Dynamics of inhomogeneous one-dimensional
coupled map lattices,''} {\em http://arxiv.org/abs/nlin.CD/0409014}.

Sano, T., Sawanobori, T. \& Adaniya, H. [1978] {``Mechanism of rhythm
determination among pacemaker cells of the mammalian sinus node,''} {\em Am. J.
Physiol.} {\bf 235}, H379--H384.

Shibata, T. \& Kaneko, K. [1998] {``Collective chaos,''} {\em Phys. Rev. Lett.}
{\bf 81}(19), 4116--4119.

Vasil'ev, K. A., Loskutov, A., Rybalko, S. D. \& Udin, D. N. [2000] {``Model of a
spatially inhomogeneous one-dimensional active medium,''} {\em Theor. Math.
Phys.} {\bf 124}(3), 1286--1297.

Wiener, N. \& Rosenblueth, A. [1946] {``Conduction of impulses in cardiac
muscle,''} {\em Arch. Inst. Cardiol. Mex.} {\bf 16}, 205--265.

Willeboordse, F. H. \& Kaneko, K. [1994] {``Bifurcations and spatial chaos in an
open flow model,''} {\em Phys. Rev. Lett.} {\bf 73}(4), 533--536.

Winfree, A. T. [2000] {\em The Geometry of Biological Time} (Springer-Verlag, New
York), 2nd ed.

\newpage

\newpage

\newpage

\newpage

\newpage

\newpage

\end{document}